\documentclass{iau}
\usepackage{graphicx}

\title[IAU Symposium 302.~~The origin of discrete absorption components] 
{Investigating the origin of cyclical spectral variations in hot, massive stars}

\author[Alexandre David-Uraz et al.]   
{Alexandre David-Uraz$^1$,
 Gregg A. Wade$^2$, V\'{e}ronique Petit$^3$ \and Asif ud-Doula$^4$}

\affiliation{$^1$Queen's University, Canada, $^2$RMC, Canada, $^3$University of Delaware, USA, \\$^4$Penn State University, USA}

\pubyear{2013}
\volume{302}  
\pagerange{119--126}
\jname{Magnetic Fields Throughout Stellar Evolution}
\editors{A.C. Editor, B.D. Editor \& C.E. Editor, eds.}
\begin{document}

\maketitle

\begin{abstract}
OB stars are known to exhibit various types of wind variability, as detected in their ultraviolet spectra,
amongst which are the ubiquitous discrete absorption components (DACs). These features have
been associated with large-scale azimuthal structures extending from the base of the wind to its outer regions: corotating
interaction regions (CIRs). There are several competing hypotheses as to which physical processes may perturb the star's
surface and generate CIRs, including magnetic fields and non radial pulsations (NRPs), the subjects of
this paper with a particular emphasis on the former. 
Although large-scale magnetic fields are ruled out, magnetic spots deserve further
investigation, both on the observational and theoretical fronts.
\end{abstract}

\firstsection 
\section{Introduction}

Despite their small numbers, massive stars are known to play an important role in galactic ecology,
enriching the interstellar medium with heavy elements processed in their cores, shaping their environment
via their fast, dense winds as well as by their spectacular deaths as supernovae. However, their impressive outflows
do not only influence their surroundings, but also significantly affect the evolution of massive stars themselves.

While important strides have been taken towards understanding these radiatively-driven, supersonic winds
(\cite[Castor et al. 1975]{CAK75}), there
still remains a number of pressing questions. For instance, for over 20 years, the community has struggled to
explain the presence of cyclical variations in wind-sensitive, UV resonance lines. The most common
manifestation of this type of variability appears as ``discrete absorption components" 
(DACs, e.g. \cite[Kaper et al. 1996]{Kaper_etal96}), which migrate
from zero velocity to approximately terminal velocity over a relatively well-defined timescale, and are thus thought to be linked to structures
extending from the base of the photosphere all the way to the outer regions of the wind.

These DACs also possess other revealing properties. Indeed, \cite[Prinja (1988)]{Prinja88} showed an apparent correlation between
the projected rotational velocities of stars and the periods of their DACs, suggesting that these variations are rotationally
modulated. Furthermore, single observations of over 200 O stars almost all show the presence of a narrow absorption component (NAC)
at more or less the terminal velocity
(\cite[Howarth \& Prinja 1989]{HowarthPrinja89}). These narrow features are believed to be snapshots of DACs, leading to the conclusion that the
DAC phenomenon is ubiquitous among hot, massive stars.

\cite[Cranmer \& Owocki (1996)]{CranmerOwocki96} provided a model to understand how these structures are generated. By introducing ad hoc
photospheric perturbations on the star (bright spots), they were able to reproduce corotating interaction regions (CIRs),
which lead to DAC-like features in synthetic UV spectra. This view has not changed dramatically since and is still considered
to be the canonical way of producing DACs. However, the nature of these perturbations is unknown. Because of the ubiquitous quality of DACs,
gaining insight on the physical process at their origin should provide meaningful information about all OB stars.

\section{Potential physical causes}

There are two main probable causes of DACs identified in the literature: magnetic fields and non radial pulsations. Furthermore,
we will distinguish between large-scale and small-scale magnetic fields. The following subsections explore each possibility in further detail.

 

\subsection{Non radial pulsations}

Non radial pulsations are known to exist in a number of well-studied DAC stars (\cite[de Jong et al. 1999]{deJong_etal99}),
but it has been difficult to draw any links between both phenomena. Observationally, these pulsations are relatively easy to detect.
Theoretically, there are some inconsistencies. Pulsations generally 
create a pattern of alternating bright and dark regions on the surface of a star. However, \cite[Cranmer \& Owocki (1996)]{CranmerOwocki96}
found that only bright spots reproduce the expected pattern. Additionally, NRP periods are typically
on a timescale of a few hours, while DACs have periods of the order of a few days. Nevertheless, it has been postulated that a superposition
of modes could create variations consistent with DAC recurrence timescales (\cite[de Jong et al. 1999]{deJong_etal99}). 
An in-depth analysis of the merits and likelihood of this hypothesis
goes beyond the scope of this paper, which focuses primarily on the magnetic properties of DAC stars; therefore, such a discussion will be
left for a later study.

\subsection{Large-scale magnetic fields}

Magnetic fields in massive stars are rare and are usually organized and dipolar in nature
(\cite[Wade \& the MiMeS Collaboration 2010]{WadeMimes10}). Consequently, they are believed to be of fossil origin.
Magnetism has been proposed a number of times as a possible cause for
DACs (e.g. \cite[Kaper \& Henrichs 1994]{KaperHenrichs94}) and given the fact that DACs generally come in pairs
(\cite[Kaper et al. 1996]{Kaper_etal96}), dipolar fields seem like a reasonable origin, albeit a challenging one. Indeed, if the small
rate of detection of magnetic fields in OB stars (7 \%, Wade et al., these proceedings) does not already raise red flags, 
the apparent inconsistency of their stable configurations
with the cyclical (rather than periodic) nature of DAC recurrence is troublesome. 

Nonetheless, \cite[David-Uraz et al., \textit{in prep.}]{DU13} have
studied the magnetic properties of 14 OB stars known or believed to have DACs. Using high-resolution spectropolarimetry in the Stokes I and V
parameters from ESPaDOnS (Canada-France-Hawaii Telescope) and NARVAL (T\'{e}lescope Bernard Lyot), they have applied a multi-line signal-enhancing
technique (Least-Squares Deconvolution, or LSD, \cite[Donati et al. 1997]{Donati_etal97}) as well as nightly-averaging to obtain high signal to noise ratio
V profiles, which were then used to try to detect Zeeman signatures. 
Assuming the oblique rotator model (\cite[Stibbs 1950]{Stibbs50}), two magnetic diagnostics are measured from the data. Longitudinal
magnetic field measurements are performed using the first-order moment of the V profiles (\cite[Wade et al. 2000]{Wade_etal00}), and 
dipolar field strengths are computed by fitting each profile and performing a Bayesian inference-based modeling, as described by
\cite[Petit \& Wade (2012)]{PetitWade12}.

\begin{figure}[htb]
\begin{center}
 \includegraphics[width=4.4in]{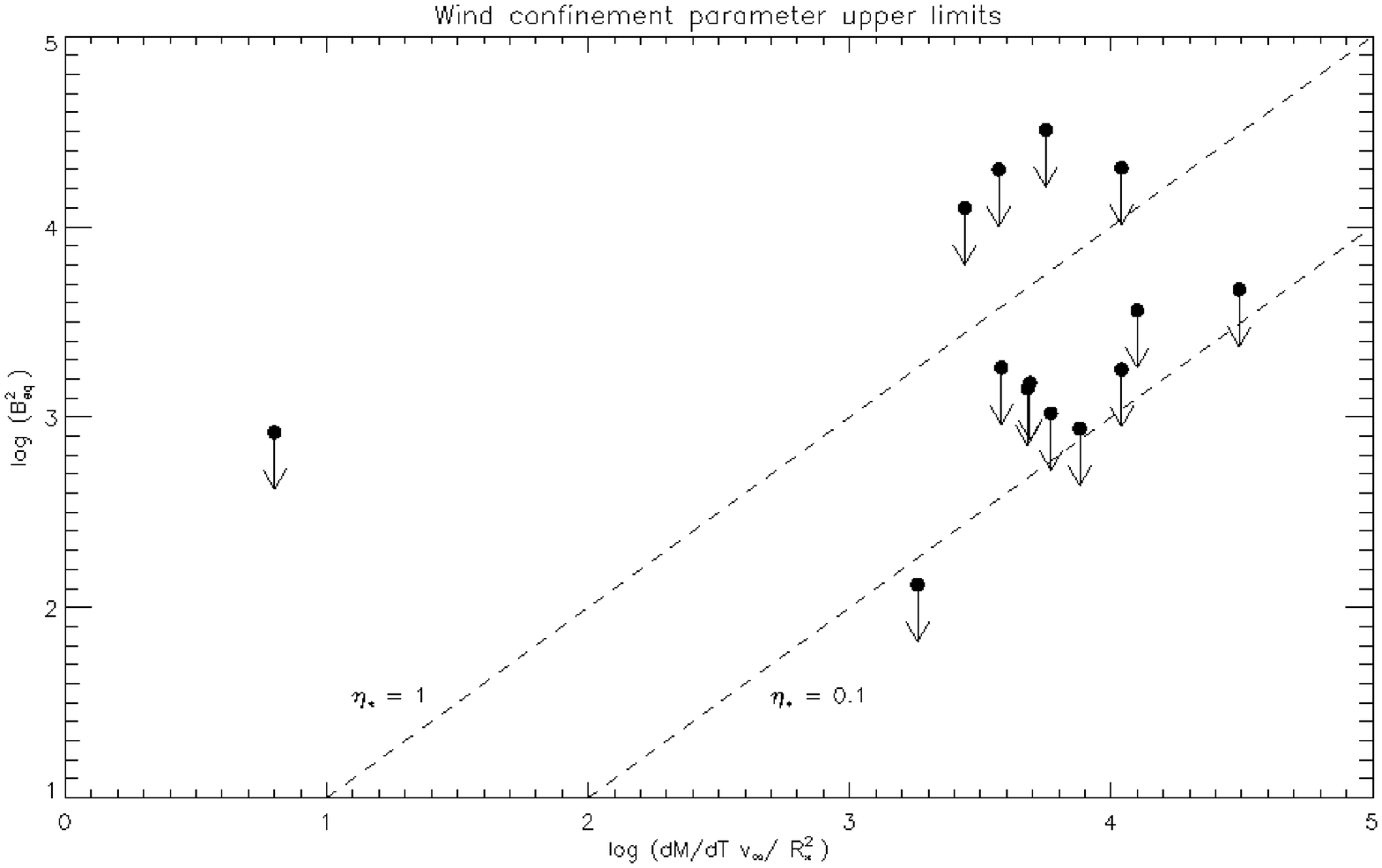} 
 \caption{Upper limits on the wind confinement parameter value for 14 well-known DAC stars. The horizontal axis corresponds loosely to the
wind kinetic energy density, while the vertical axis is proportional to the magnetic energy density. The dashed lines correspond to wind confinement
parameter values of 1 and 0.1, which are considered as important thresholds (see section 2).}
   \label{fig1}
\end{center}
\end{figure}

The results are unequivocal: no dipolar magnetic fields are detected. All derived values are consistent with a null magnetic field. Furthermore, 
extremely tight constraints are obtained (the best constraints on both nightly longitudinal field error bars, 4 G, and dipolar field strength, 23 G,
are obtained for the sharp-lined O dwarf 10 Lac). Two interaction mechanisms between the magnetic field and the
wind are considered: magnetic wind confinement, and magnetically-induced surface brightness enhancements. In the first case, the magnetic field entraps material by
restricting the outflow when it is perpendicular to field lines (i.e. at the equator). Even when the material is not completely confined, this 
type of mechanism dynamically influences the outflow. This interaction was described by \cite[ud-Doula \& Owocki (2002)]{udDoulaOwocki02}, who define
the ``wind-confinement parameter" essentially as the ratio of the magnetic energy density and the wind kinetic energy density:
$\eta_{*} = ({B_{eq}}^2 {R_{*}}^2)/(\dot{M} v_{\infty})$,
where $B_{eq}$ corresponds to the strength of the magnetic field at the equator (which equals half of the dipolar field strength),
$R_{*}$ is the stellar radius, $\dot{M}$ is the mass-loss rate and $v_{\infty}$ is the terminal velocity of the wind. They showed that above a
threshold of $\eta_{*} = 1$, the wind is magnetically confined, whereas it is dynamically influenced all the way down to $\eta_{*} = 0.1$. 
Figure~\ref{fig1} shows the upper limits derived from the Bayesian analysis. Less than half (5) of the points have upper limits above $\eta_{*} = 1$, 
slightly more than half (8) are between both thresholds (with a few of them quite close to the lower one), and one point (10 Lac) is under $\eta_{*} = 0.1$.
Clearly, most of these stars do not have magnetically confined winds, therfore a large portion
of their winds could not be dynamically influenced by globally-organized dipolar magnetic fields, should they exist. 

%

For the bright spot mechanism, the idea is that the magnetic pressure within a flux tube would decrease the gas pressure (as compared to a point
on the surface with no magnetic flux), therefore creating ``wells" which probe deeper into the star, which result in hot spots (given the temperature
gradient in hot star atmospheres). The luminous flux enhancement can be written as follows: $F'/F = 1 + (3 \kappa B^{2})/(32 \pi g)$,
where $\kappa$ is the mean Rosseland opacity, $B$ is the magnetic field strength and $g$ is the gravitational acceleration. Finally,
taking typical values for O dwarfs ($\kappa \sim 1$ and $\log g = 4.0$), it is estimated that a 400 G magnetic field is required to produce
a 50\% brightness enhancement (this corresponds to the value used by \cite[Cranmer \& Owocki 1996]{CranmerOwocki96}), which is much greater than all the dipolar
field strength upper limits obtained for the sample. In conclusion, combined with the tight constraints obtained observationally, 
both mechanisms fail to create the required conditions to generate DACs.

%

\subsection{Small-scale magnetic fields}

On the other hand, the case for magnetism in general is not settled. Indeed, \cite[Cantiello \& Braithwaite (2011)]{CantielloBraithwaite11}
argue that the iron opacity bump at 150 kK leads to a sub-surface convection zone in the most massive stars, resulting in potentially
measurable small-scale surface magnetic fields (or spots). Unfortunately, the characteristics of these fields, as inferred from theory, 
are not currently very well constrained, leading
to significant modeling challenges. While these spots would be very hard to detect (especially if they are distributed
randomly, \cite[Kochukhov \& Sudnik 2013]{KochukhovSudnik13}), they present an attractive alternative to other field configurations. Indeed, spots
on the Sun for instance vary with time; transient spots seem consistent with the cyclical recurrence of DACs. Furthermore, this fits into
a picture where DACs are modulated by rotation.

\section{Future work}

As mentioned earlier, it has been clearly demonstrated that dipolar fields are not responsible for the DAC phenomenon. Therefore, to further test the magnetic
hypothesis, more attention now has to be given to small-scale fields. On the observational side, deeper magnetic measurements can be performed on a limited
number of well-suited stars (bright, massive, low projected rotational velocity), and then analyzed using a Bayesian inference technique similar
to that used for dipolar fields (only, in this case, using a parametrized spot model under a certain number of assumptions). High-precision photometry
has also been suggested as a means of detecting CIRs (e.g. \cite[Chen\'{e} et al. 2011]{Chene_etal11}, \cite[David-Uraz et al. 2012]{DU12}, etc.) and
deserves to be looked into.

Further numerical simulations should be conducted to show whether magnetic spots can actually produce DAC-like signatures, and if so, what the relevant
parameter space is. Non radial pulsations should also be studied more systematically. Finally, if all else fails, it might be necessary to think outside
the box, for instance, challenging the idea that DACs are rotationally modulated, and trying to find another mechanism to create them. In the end,
that might just be what it takes to solve this long-standing problem.

\end{document}